\documentclass[conference]{IEEEtran}
\IEEEoverridecommandlockouts
\usepackage{cite}
\usepackage{comment}
\usepackage{amsmath,amssymb,amsfonts}
\usepackage{algorithmic}
\usepackage{graphicx}
\usepackage{textcomp}
\usepackage{xcolor}
\usepackage{booktabs}
\usepackage[hidelinks]{hyperref}
\usepackage{url}
\usepackage{makecell}

\def\BibTeX{{\rm B\kern-.05em{\sc i\kern-.025em b}\kern-.08em
    T\kern-.1667em\lower.7ex\hbox{E}\kern-.125emX}}
\begin{document}

\title{SAMAY: System for Acoustic Measurement and Analysis \\
}

\author{
\IEEEauthorblockN{Adheep Arya G R}
\IEEEauthorblockA{\textit{IoT Group} \\
\textit{C-DAC Bangalore}\\
Bangalore, India \\
adheepgr@cdac.in}

\\
\IEEEauthorblockN{Ruchi Juyal}
\IEEEauthorblockA{\textit{IoT Group} \\
\textit{C-DAC Bangalore}\\
Bangalore, India \\
ruchijuyal@cdac.in}

\and
\IEEEauthorblockN{Vaibhav Pratap Singh   }
\IEEEauthorblockA{\textit{IoT Group} \\
\textit{C-DAC Bangalore}\\
Bangalore, India \\
svaibhav@cdac.in}
\\
\IEEEauthorblockN{Sangit Saha}
\IEEEauthorblockA{\textit{IoT Group} \\
\textit{C-DAC Bangalore}\\
Bangalore, India \\
sangitsaha@cdac.in}
\and
\IEEEauthorblockN{Mayank Kumar}
\IEEEauthorblockA{\textit{IoT Group} \\
\textit{C-DAC Bangalore}\\
Bangalore, India \\
kmayank@cdac.in}
\\
\IEEEauthorblockN{Kaushik Nanda}
\IEEEauthorblockA{\textit{IoT Group} \\
\textit{C-DAC Bangalore}\\
Bangalore, India \\
nandak@cdac.in}
\and
\IEEEauthorblockN{Niyathi Shenoy}
\IEEEauthorblockA{\textit{IoT Group} \\
\textit{C-DAC Bangalore}\\
Bangalore, India \\
niyathishenoy@cdac.in}

\\
\IEEEauthorblockN{Hari Babu Pasupuleti}
\IEEEauthorblockA{\textit{IoT Group} \\
\textit{C-DAC Bangalore}\\
Bangalore, India \\
hari@cdac.in}

\and
\IEEEauthorblockN{Tejas Suryawanshi}
\IEEEauthorblockA{\textit{IoT Group} \\
\textit{C-DAC Bangalore}\\
Bangalore, India \\
tejassunils@cdac.in}
\\\IEEEauthorblockN{S D Sudarsan}
\IEEEauthorblockA{\textit{Executive Director} \\
\textit{C-DAC Bangalore}\\
Bangalore, India \\
sds@cdac.in}
}
\maketitle

\begin{abstract}

This paper describes an automatic bird call recording system called SAMAY, which is developed to study bird species by creating a database of large amounts of bird acoustic data. By analysing the recorded bird call data, the system can also be used for automatic classification of bird species, monitoring bird populations and analysing the impact of environmental changes. The system is driven through a powerful STM32F407 series microcontroller, supports 4 microphones, is equipped with 128 GB of storage capacity, and is powered by a 10400 mAh battery pack interfaced with a solar charger. In addition, the device is user-configurable over USB and Wi-Fi during runtime, ensuring user-friendly operation during field deployment. 

\end{abstract}


\section{Introduction}

Birds play a vital role in maintaining the stability and functionality of the ecosystem.  In addition, birds are highly responsive to environmental changes, making them excellent indicators of the ecosystem \cite{knight}.  However, biodiversity studies have highlighted a significant global decline in bird populations, with many species becoming increasingly vulnerable due to habitat destruction, climate change, and pollution. These trends emphasise the urgent need for thorough monitoring to develop effective conservation strategies to protect avian species and their habitats. It is crucial to collect and process recordings over extended periods to completely study the behavioural patterns of different bird species with changing environmental conditions. 

Various animal species, particularly birds, frequently produce sounds, each carrying a distinct "acoustic fingerprint" unique to its species. As a result, birds can be identified based on their calls, making acoustic classification an effective method for studying avian populations, understanding species demographics, and studying behavioural patterns. Acoustic monitoring has emerged as a nonintrusive approach to recording avian sounds. By analysing these recordings, ornithologists can assess bird diversity, behavioural adaptations, and seasonal population changes, and identify endangered species. To achieve this, autonomous recording units (ARUs) are essential, as they offer unbiased and consistent monitoring, especially for species that actively avoid human presence.

The developed 'System for Acoustic Measurement and Analysis' (SAMAY) is an automated bird call recording system used for passive acoustic monitoring. It is designed to collect long-term acoustic data to study different bird species and their habitat. The proposed system is well-suited for long-term deployment because of its large storage capacity and solar chargeable battery operation.

\section{Related Work and Motivation}

Verma et al. have developed, an IoT-based acoustic monitoring system designed to support bird conservation through automated audio detection and remote data transmission. The device is equipped with a MEMS microphone for high-quality audio capture, a 64 GB SD card for local storage—ensuring reliable long-term data retention and a 4G modem for sending processed audio data to the cloud using the lightweight MQTT protocol. This combination helps reduce both operational costs and dependency on continuous network connectivity. In addition, the system incorporates a rapid target species detection algorithm (DA) \cite{AviEar}.

Kiarie at al. have developed a Raspberry Pi-based recording system for acoustic bird monitoring. The recording system also integrates a solar power supply \cite{RPI}. Latorre et al. have developed a low-power embedded audio recording system, which utilizes MEMS microphones. To minimize storage and power consumption, audio compression algorithm, Adaptive Differential Pulse-Code Modulation (ADPCM), is also implemented \cite{MEMS}.

Magno et al. and Tiong et al. developed a compact, low-power sensor node specifically designed to be worn by small birds for continuous monitoring of vocal and physiological activity. The system, which communicates with a remote host via Bluetooth Low Energy (BLE), enables seamless data transfer while minimizing energy consumption. One of the standout features of this solution is its ability to stream both compressed and uncompressed audio in real time, along with temperature measurements, offering flexibility for different research needs. This approach supports high-fidelity acoustic monitoring and opens new possibilities for studying avian behavior in controlled environments \cite{Blue}\cite{tiong2023embedded}.

Beason et al. developed AURITA, an affordable autonomous recording device to monitor bird and bat species. The system is capable of recording both audible and ultrasonic frequencies and supports continuous 24-hour recording with a sampling rate of 44.1 kHz, making it a versatile tool for bioacoustics and soundscape research \cite{beason2019aurita}.

\begin{figure}[!t]
\centerline{\includegraphics[scale=0.38]{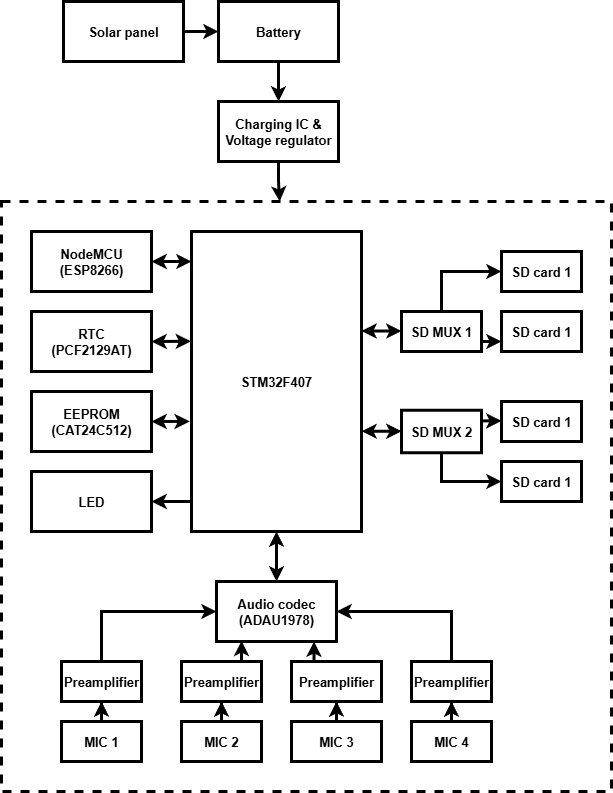}}
\caption{Block Diagram of SAMAY Board.}
\label{fig:SAMAY_Block}
\end{figure}
Hill et al. developed AudioMoth, a low-cost and compact acoustic monitoring device for biodiversity research \cite{hill2019audiomoth}. The system is built on a printed circuit board (PCB), a microcontroller, and a MEMS microphone, offering a lightweight and energy-efficient solution for long-term environmental sound recording \cite{hill2019audiomoth}.

A few commerical solutions are also available in the market. SongMeter4 is a professional recorder unit known for its durability and superior recording quality in difficult field conditions. It delivers cleaner audio with a higher signal-to-noise ratio and features like GPS, making it more robust and reliable for long field deployments\cite{wildlifeacoustics2024}. Swift is a lightweight, cost effective recorder unit supporting to sampling up to 96 kHz (16-bit) and is considered suitable for long deployments with minimal maintenance\cite{cornell2023}. Arbimon uses an Android device preloaded with Arbimon Touch, which allows users to configure names ,recording schedules, file formats and microphone gain. Designed for field use, it balances portability, durability, and user-friendly operation for biodiversity monitoring\cite{sieve2015}.

Some recent studies, such as Sanchez et al. and Changapur et al., have also explored the integration of machine learning techniques for bioacoustic classification\cite{article}\cite{Bio_sensor}\cite{ML_1_bng}. 

Although significant progress has been made in bioacoustic monitoring, there still remains a need to design robust, low-power, high-storage solutions that incorporate indigenous designs, efficient storage management, rechargability and on-board advanced processing capabilities to improve bird monitoring in diverse environments. In this paper we introduce such an indigenous system, SAMAY, detailing the system architecture, operational flow and functional test results. 

\section{System Architecture}

\begin{figure}[!t]
\centerline{\includegraphics[scale=0.12]{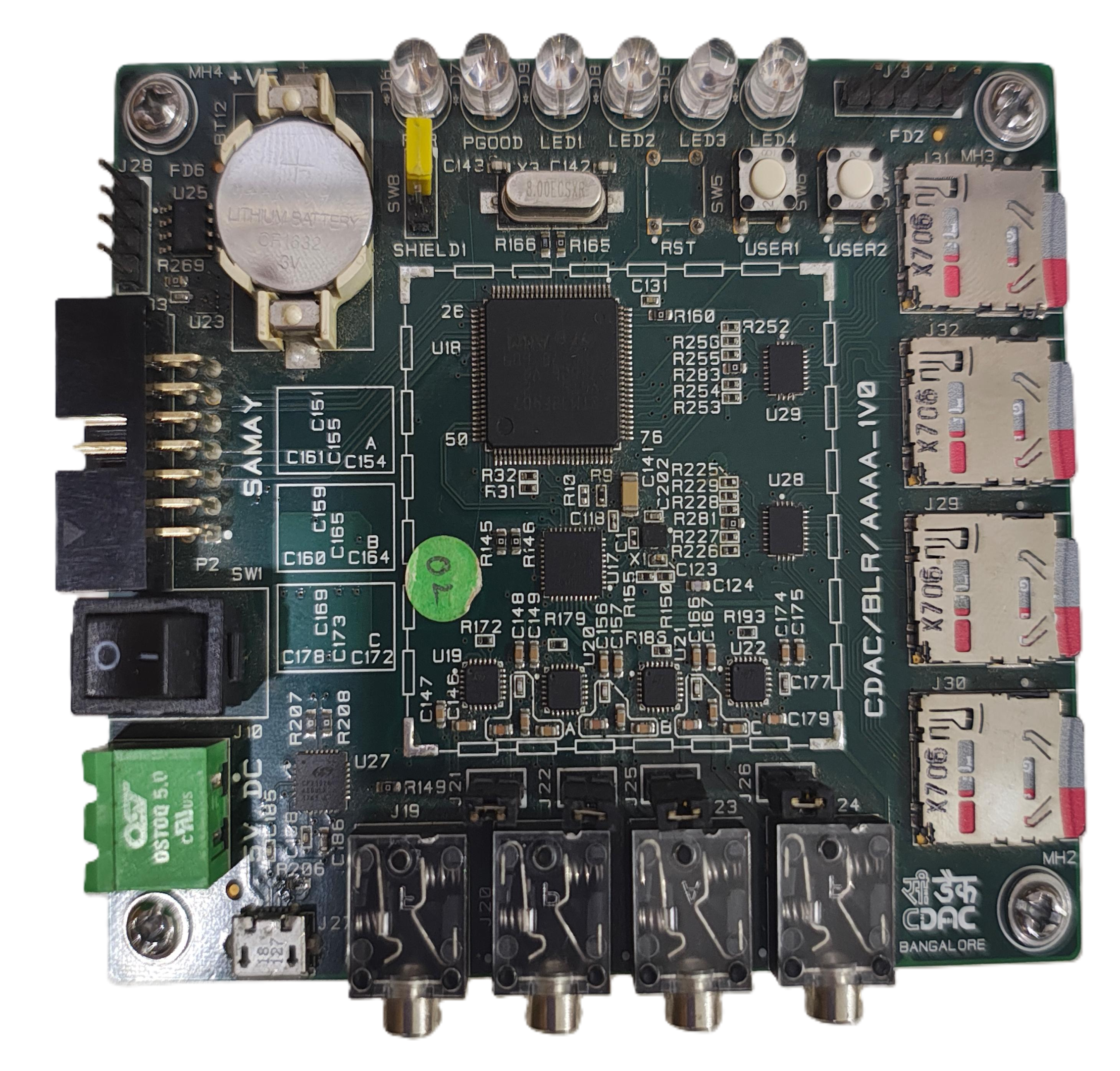}}
\caption{SAMAY Indigenous PCB Board}
\label{fig:SAMAY_1}
\end{figure}

\begin{figure}[!b]
\centerline{\includegraphics[scale=0.9]{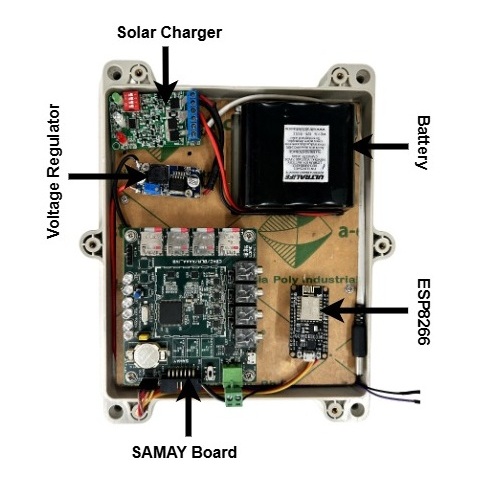}}
\caption{SAMAY internal overview.}
\label{fig:SAMAY overview}
\end{figure}

SAMAY is a battery-operated electronic device designed for field deployment to capture the activity of bird species using acoustic data. The block diagram and assembled internal PCB board of the proposed SAMAY system, illustrating its various subunits and interconnections, are presented in Fig. \ref{fig:SAMAY_Block} and  Fig. \ref{fig:SAMAY_1}, respectively. Table \ref{table:hard} gives an overview of the components used in the SAMAY board, and Fig. \ref{fig:SAMAY overview} gives an overview of the internal circuit.  
The block diagram comprises three main hardware blocks:

1) system on chip (SoC).

2) peripheral components.

3) power control and charging modules.

\begin{table}[!b]
    \centering
    \caption{Major Components of SAMAY}
    \begin{tabular}{|l|l|l|}   
        \hline
        \textbf{Components}         & \textbf{Functions}                     & \textbf{Manufacturer}       \\ 
        \hline  
        STM32F407                   & System on chip                         & ST-Microelectronics          \\ \hline
        ADAU1978                    & Audio codec                            & Analog Devices               \\ \hline
        PCF2129AT                   & Real-time-clock                        & NXP                          \\ \hline
        CAT24C512                   & User-modifiable memory                 & Onsemi                       \\ \hline
        TS472                       & Preamplifier                           & ST-Microelectronics          \\ \hline
        FSSD06                      & Secure digital multiplexer             & Fairchild                    \\ \hline 
        \makecell[l]{ESP8266}       & User device configuration        & Amica     \\ \hline

        LT3652                      & Charging IC                            &Analog Devices                \\ \hline
    \end{tabular}
    \label{table:hard}
\end{table}

\subsection{System on Chip}
SAMAY's central component is the STM32F407 microcontroller ($\mu C$)\cite{STM32F407}, a powerful 32-bit Arm® Cortex®-M4 CPU featuring a floating point unit (FPU). It operates at speeds of up to 168 MHz, delivering 210 Dhrystone Million Instructions Per Second (DMIPS).  The microcontroller is ideal for embedded applications that need effective data processing and storage because it has 192 KB of SRAM and 1 MB of Flash memory.

This chip supports multiple low-power modes and offers a range of communication interfaces, including SPI, I2C, and UART, along with SDIO for high-speed data storage. The SAMAY system, is responsible for:

(1). Real-Time Clock (RTC) and EEPROM management

(2). Configuring the audio codec via the I2C protocol

(3). Writing data to an SD card using SDIO with a 4-bit wide bus

(4). Receiving and storing digital audio signals from audio codec via the I2S protocol

\subsection{Peripheral Components}
\subsubsection{Audio codec}
The ADAU1978 \cite{ADAU1978} is a high-performance, low-power analog-to-digital converter (ADC) designed for audio applications. It features four differential input channels with 24-bit resolution and supports sample rates ranging from 8 kHz to 192 kHz, ensuring high-quality audio capture. Additionally, it incorporates a programmable gain amplifier (PGA) with adjustable gain from 0 dB to 60 dB in 0.375 dB steps. The ADC also includes an integrated high-pass filter to eliminate DC offsets and operates with low latency, making it suitable for real-time audio processing.
In this application, the ADAU1978 \cite{ADAU1978} processes analog signals received from the preamplifier and transmits the digitized data to the $\mu C$ using the I2S protocol.

\subsubsection{ESP8266}
 The ESP8266 is a low-cost, Wi-Fi-enabled microcontroller module based on the ESP8266 chip. It can be used for collecting device configuration from a mobile application and transmitting it to the $\mu C$. The $\mu C$ then uses this data to configure the device accordingly \cite{ESP8266}.

\subsubsection{RTC}
The PCF2129 \cite{PCF2129AT} is a CMOS Real-Time Clock (RTC) with an integrated Temperature Compensated Crystal Oscillator (TCXO) and a 32.768 kHz quartz crystal, ensuring high accuracy and low power consumption. It supports both I2C and SPI communication protocols. In SAMAY, the PCF2129 serves as the primary timer for alarm interrupts. It is configured to generate an alarm at a user-defined time, which triggers the start of recording at the specified moment. 

\subsubsection{EEPROM}
The CAT24C512 \cite{CAT24C512} is a 512 KB Serial EEPROM internally organized into 65,536 words of 8 bits each. This EEPROM is used to store configuration packets received from the user. When the device powers on, it reads the stored configuration and restores previous settings.

\subsubsection{Preamp}
The TS472 \cite{TS472} is a differential-input microphone preamplifier optimised for high-performance PDA and notebook audio systems. This device features an adjustable gain from 0 to 40 dB with an excellent power supply and common-mode rejection ratios.
In this application, the TS472 \cite{TS472} preamplifier is used to amplify audio signals received from the microphone, ensuring clear and strong signals for further processing.

\subsubsection{SD MUX}
In this system, the FSSD06 \cite{FSSD06} SD multiplexer is employed to interface and manage up to four SD cards (with two SD slots per multiplexer), enabling dynamic switching between multiple storage devices without requiring physical intervention. This allows for efficient data logging, storage expansion, and redundancy, which is especially useful in long-term or mission-critical applications such as environmental monitoring, data backup systems, or embedded IoT devices. The multiplexer ensures seamless communication between the host microcontroller and the selected SD card using standard SDIO protocols, while also helping prevent bus contention and signal interference when multiple cards are present. 

\subsubsection{Power Control and Charging Module}
The LT3652 \cite{LT3652} is a compact step-down battery charger designed to handle input voltages ranging from 4.95V to 32V, and it can deliver up to 2A of charge current using constant-current/constant-voltage (CC/CV) charging. It includes useful features like input voltage regulation, automatic charge termination, auto-restart, and an adjustable safety timer, which make it especially well-suited for solar-powered systems. In this setup, the IC is used to charge a 7.2V, 10,400mAh Li-ion battery pack from a 10W solar panel, providing reliable charging with built-in protection against overcharging, deep discharge, and short circuits.

\subsection{Enclosure}

The SAMAY system is enclosed in an IP65 dustproof, watertight hard case with dimensions of 250 × 185 × 85 mm, designed for rugged outdoor environments. The enclosure can be easily attached to poles or trees for convenient deployment.

\section{Operational Details}

\begin{figure}[htbp]
\centerline{\includegraphics[scale=0.42]{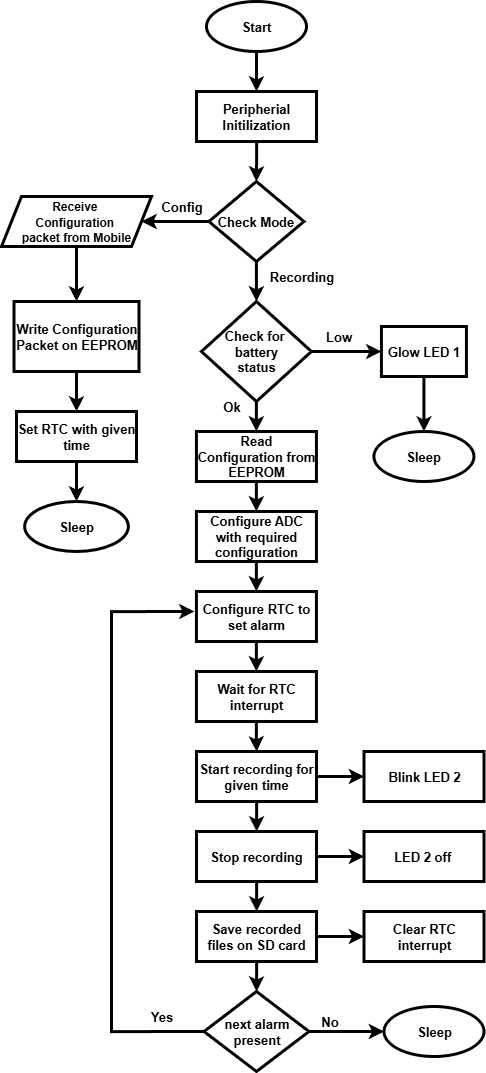}}
\caption{SAMAY operational flowchart.}
\label{fig:SAMAY Audio recording flowchart}
\end{figure}

Fig. \ref{fig:SAMAY Audio recording flowchart} shows the operational flow diagram of the SAMAY device. Upon boot-up, the device initializes all connected peripherals, including the audio codec, RTC, and EEPROM etc. It then determines the mode of operation, which can be either Configuration mode or Recording mode. The mode can be set through a jumper at every restart, and over can be done via the mobile app. 

\subsection{Configuration Mode}
In Configuration mode, the device waits for the user to send device configuration data through a mobile app using a specific packet ID. The configuration can be done over Wi-Fi or USB. Configuration over Wi-Fi is sent to the $\mu C$ via ESP8266, through a UART interrupt. $\mu C$ processes the received data based on its ID and stores in EEPROM to retain the configuration settings. The device configuration can be changed during non-recording ON time.   

\subsection{Recording Mode}
In Recording Mode, the device first checks the battery status, and if the battery level is sufficient, it proceeds to read the stored configuration from the EEPROM and sets the RTC alarm accordingly. The device then enters a low-power sleep state, awakening only when the RTC interrupt occurs to start recording. The recording process begins by initializing the I2S DMA, which collects data from the ADC and stores it in a 4096-byte local buffer. As soon as the first half of the buffer is filled, an interrupt is triggered to write these data to a .wav file, while the other half continues to fill via DMA. Once the full buffer is filled, another interrupt occurs, writing the second half to the .wav file. This double buffering technique prevents data loss due to the continuous nature of DMA transfers, which could otherwise overwrite buffer contents before they are saved.  
The recorded data is saved as a .wav file. After the defined recording duration, the system clears the RTC interrupt, stops the I2S DMA, and saves the .wav file to the SD card. The SD card is controlled through SDIO and formatted with FATFS. The device then checks for the next scheduled alarm, and if another recording session is set, the RTC alarm is reconfigured accordingly. Since the system utilizes two I2S peripherals, two separate .wav files are recorded per session, corresponding to the data from each I2S interface.


\section{RESULTS: Functional Validation}

\begin{figure}[!t]
\centerline{\includegraphics[scale=0.11]{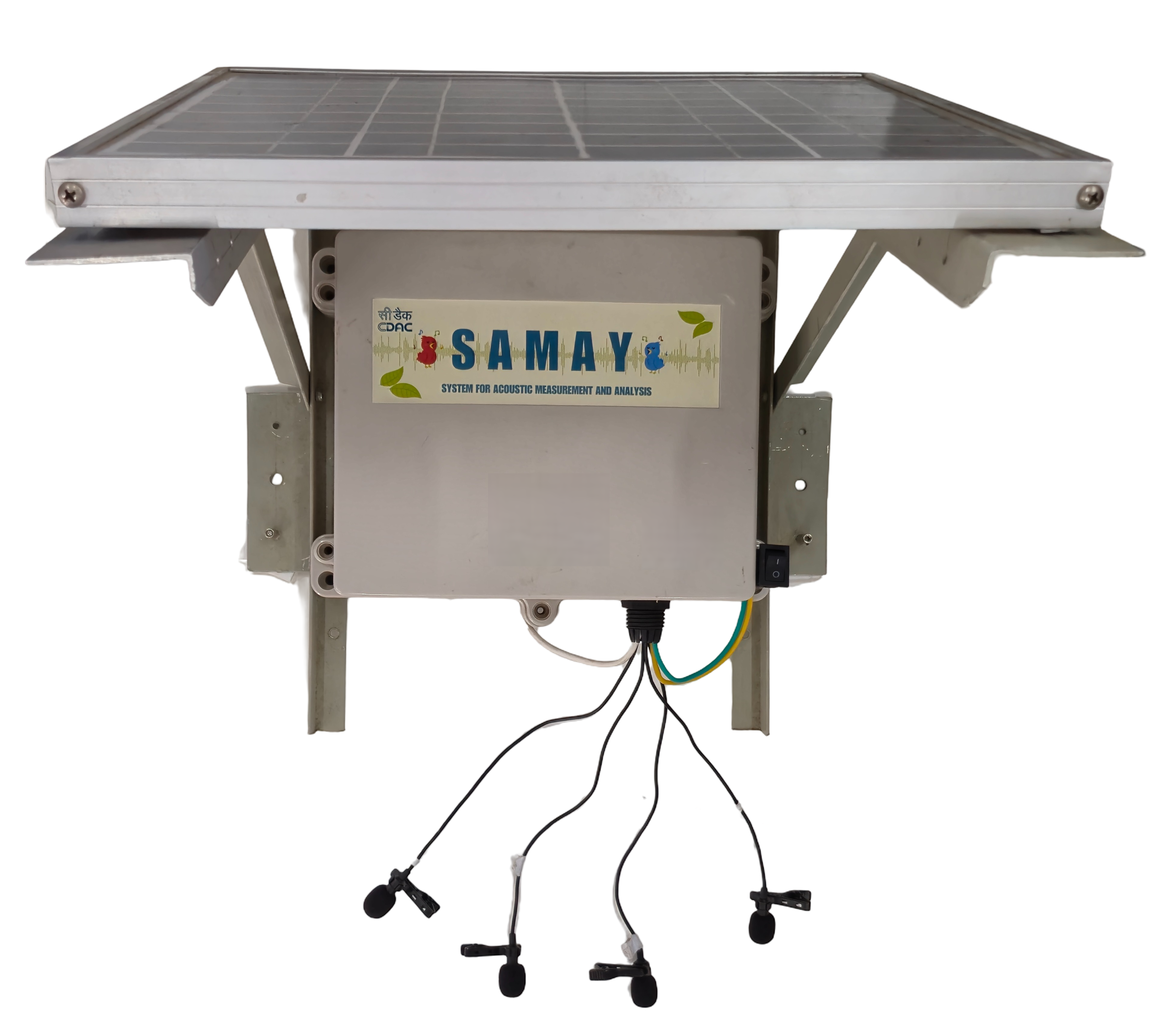}}
\caption{Enclosed SAMAY device.}
\label{fig:samay_enclosure}
\end{figure}

Fig. \ref{fig:samay_enclosure} shows the completely enclosed SAMAY bird call recording device. Based on the stored configuration in EEPROM, the system sets RTC interrupts to wake the device at the designated times and initiate recording automatically.
In addition, SAMAY is equipped with solar charging, making it well-suited for long-term bird monitoring in remote locations. 

\begin{figure}[!b]
\centerline{\includegraphics[scale=0.19]{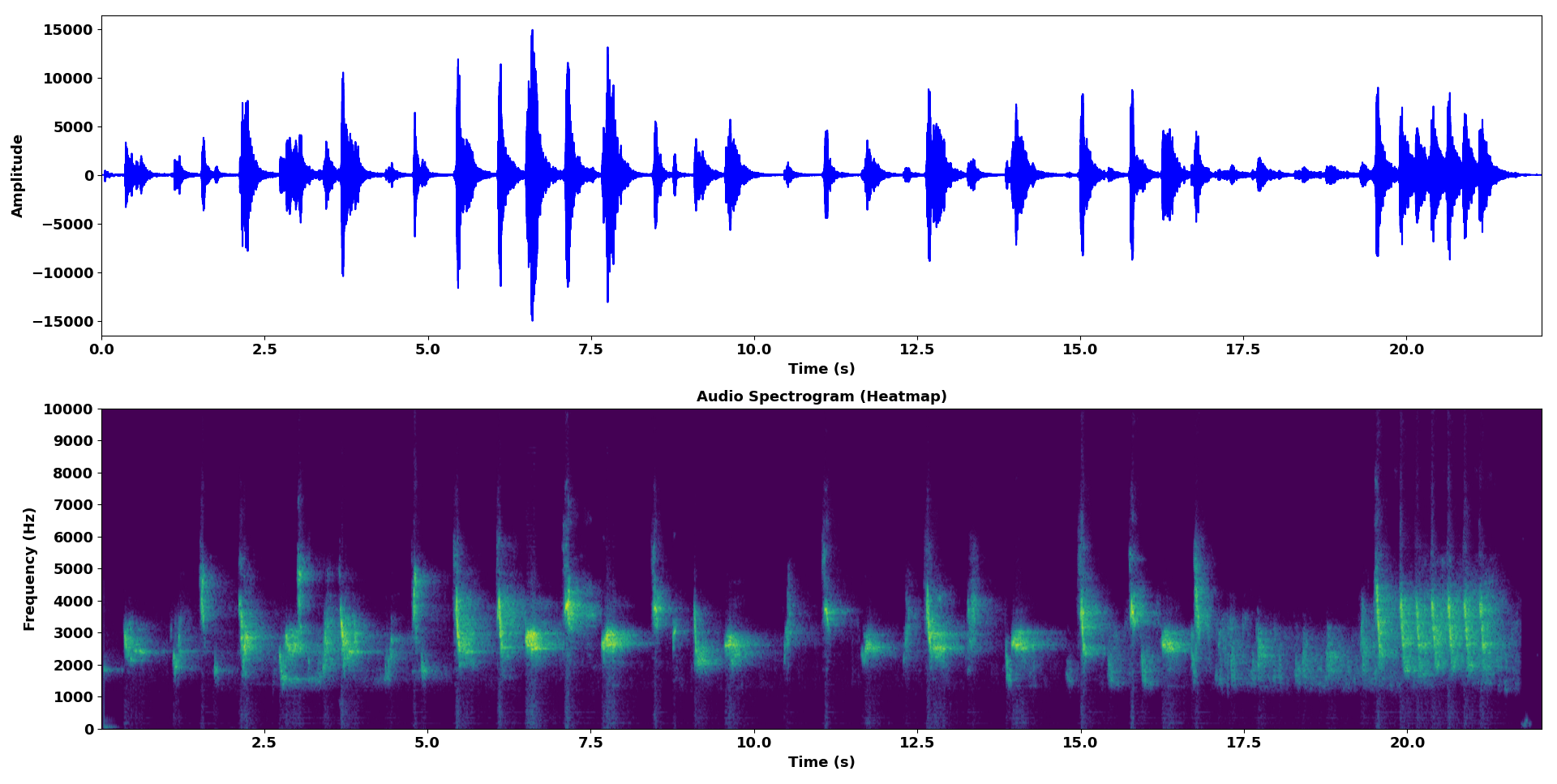}}
\caption{Waveform and spectrogram of a file recorded using SAMAY.}
\label{fig:Waveform and Spectrogram of a File Recorded using SAMAY}
\end{figure}

Fig. \ref{fig:Waveform and Spectrogram of a File Recorded using SAMAY} presents a time variation of a bird call recorded using the SAMAY device along with its spectrogram. The spectrogram reveals that higher power is concentrated in the audible frequency ranges, which is likely attributed to recorded bird call.

\subsection{SAMAY App}
\begin{figure}[!t]
\centerline{\includegraphics[scale=0.38]{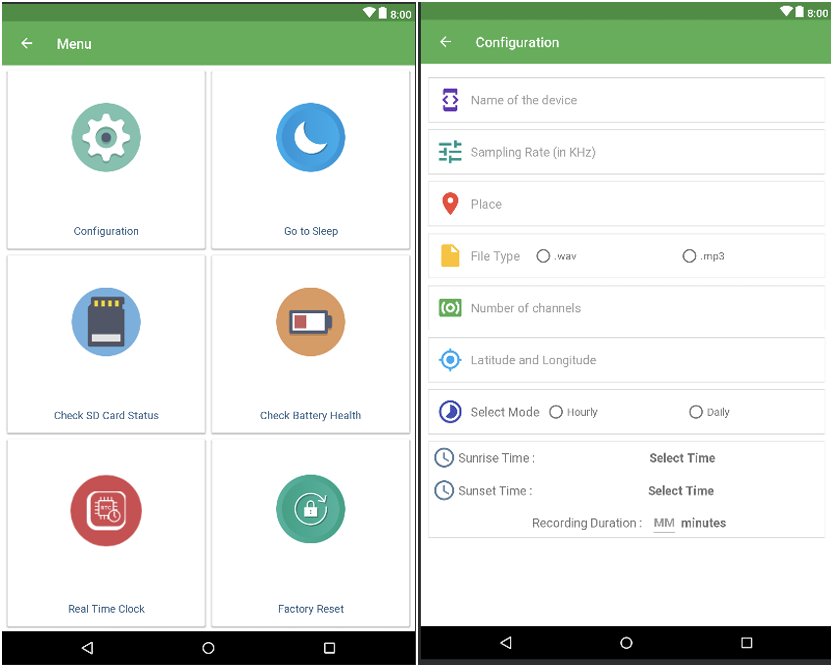}}
\caption{Mobile app interface for SAMAY.}
\label{fig:SAMAY_app}
\end{figure}
To streamline field operations, we developed a cross-platform mobile application that serves as the primary interface for configuring and managing the SAMAY audio recording device. Fig. \ref{fig:SAMAY_app} shows the mobile app interface. The app enables users to dynamically adjust critical parameters such as sampling frequency (8–192 kHz) and audio file format .wav or .mp3, ensuring compatibility with diverse ecological monitoring requirements. Operational modes include Daily and Hourly configurations. In Daily mode, the device wakes up every hour between a pre-defined sunrise and sunset time and records for 10 mins. In Hourly mode, the user sets desired wake-up times and recording duration. Other possible advanced features include battery health monitoring, SD card storage monitoring, real time clock configuration, factory reset and manual forced sleep. For targeted acoustic surveys, the app supports time-bound recording schedules, allowing users to define precise start/stop times or trigger recordings during specific environmental conditions such as dawn/dusk periods for avian activity. All configurations over Wi-Fi are stored in a JSON-based profile, that can be shared across devices, allowing standardized configurations for large-scale deployments. The USB and low-latency Wi-Fi based configuration ensure reliable operation in remote areas with limited connectivity, further enhancing the adaptability of SAMAY to challenging field environments.

\subsection{Power Consumption Analysis}

The SAMAY device operates through a 7.2V, 10,400 mAh lithium-ion battery enabled with solar charging feature. We have optimised the power consumption by actively using sleep modes across the code in different operating modes. Typical power requirements of major components are listed in Table \ref{table:power}. 

\begin{table}[ht]
\caption{Power Consumption of SAMAY Components}
\centering
\begin{tabular}{|c|c|c|}
\hline
\textbf{Component} & \textbf{Max. Current (mA)} & \textbf{Rec. Voltage (V)} \\
\hline
STM32F407 MCU & 240 mA & 1.8 V-3.6 V  \\ \hline
ADAU1978 ADC & 14 mA & 3.3 V \\ \hline
CAT24C512 EEPROM & 1.8-2.5 mA & 1.8-5.5 V\\ \hline
PCF2129AT RTC & 0.70 $\mu$A & 3.3 V \\ \hline
SD Card (Write Peak) & 30-60 mA & 3.3 V  \\ \hline
ESP8266 Wi-Fi Module & 80 mA (idle) & 3.3 V \\ 

\hline
\end{tabular}
\label{table:power}
\end{table}

We operated the device at an input voltage of 7.2 V, through a power supply instead of a battery and found the typical current consumption to be 200 mA during active recording. The power consumption falls to 179 mA during the non-recording state. The device was configured for the typical configuration of a sampling rate of 48 kHz, 16-bit depth, and stereo audio recording. We expect the sleep state power consumption to decrease further with better code optimisation, as a lower average power profile will make SAMAY suitable for extended autonomous deployment with solar energy support.

\subsection{File size calculation}
SAMAY is equipped with 128 GB of onboard storage, enabling extended field deployments in remote regions. During each recording session, the device generates two stereo-channel .wav files with user-defined durations. When configured at a sampling rate of 48 kHz, 16-bit depth, and stereo audio, a 10-minute recording produces files of approximately 10.9 MB each. Thus, a single session (comprising two files) consumes around 22 MB of storage space. Given this configuration, typically the device can store up to 1,092 individual 10-minute recordings before exhausting its storage capacity. This makes SAMAY particularly well-suited for long-term, unattended bioacoustic monitoring in field conditions where data retrieval may be infrequent.

\subsection{On board audio processing capabilities}
The usage of a high-performance STM $\mu C$ on SAMAY is to keep up with typical runtime processing requirements. The STM $\mu C$ supports to optimise audio quality by dynamically filtering noise using a bandpass filter implementation. This feature has been tested to provide better-quality audio recording during post-processing and can be automated on the $\mu C$ based on the requirements of field experts. STM's processing capabilities can also be used to optimise the storage requirement by automatically removing data corresponding to no bird activities. Frames falling below a user-defined amplitude threshold can be discarded, reducing storage requirements in different environments. Users can fine-tune parameters to balance storage efficiency and signal preservation based on the research needs of bird call experts.

\section{CONCLUSION}

In conclusion, SAMAY is a device that provides a low-power, autonomous bird audio recording device developed specifically for passive acoustic monitoring applications. It features an ARM Cortex-M microcontroller equipped with a multi-channel ADC and a 128 GB storage module, which makes the system suitable for extended field deployments for efficient data acquisition. Availability of such a system enables large database creation, with assisted bird call annotation by bird call experts. This leads to an online platform towards audio-based bird identification using machine learning techniques. The availability of 4-channel audio codec shall enable the implementation of triangulation for bird localisation in the future. Based on inputs from bird call experts, SAMAY can further be easily augmented with automated noise filtering to improve audio quality and automatic removal of data corresponding to no bird activities optimzing storage space.

\section{Acknowledgment}

The authors would like to express their gratitude to SERB, Department of Science and Technology (DST), for funding this research. The authors would like to thank Prof. Padmanabhan Rajan, IIT Mandi, Prof. Robin Vijayan, IISER Tirupati, and Prof. Anil Prabhakar, IIT Madras, for their expertise in the domain, which provided valuable insights while designing SAMAY. Authors also acknowledge ChatGPT for help with framing sentences more effectively.

\bibliographystyle{IEEEtran}
\bibliography{reference1} 

\end{document}